\documentclass[5p]{elsarticle}

\usepackage{graphicx}
\usepackage{amsmath}
\usepackage{amssymb}
\usepackage{pifont}
\usepackage{color}

\biboptions{sort&compress}

\newcommand{\req}[1]{Eq.\,(\ref{#1})}

\newcommand{\beqn}{\begin{equation}}
\newcommand{\eeqn}{\end{equation}}

\newcommand{\CUDO}{{\small CUDO}}

\begin{document}
\title{Properties of Gravitationally Bound Dark Compact Ultra Dense Objects}
\date{30 December 2011} 

\author{Christopher Dietl\fnref{fn1}}
\ead{c.dietl@stud.uni-heidelberg.de}

\author{Lance Labun\corref{cor1}}
\ead{labun@physics.arizona.edu}

\author{Johann Rafelski\corref{}}
\address{Department of Physics, University of Arizona, Tucson, Arizona, 85721 USA}

\cortext[cor1]{Corresponding author}
\fntext[fn1]{Present Address: University of Heidelberg, Heidelberg, Germany}

\begin{abstract} 

We consider compact astrophysical objects formed from dark matter fermions of mass 250 GeV to 100 TeV or from massless fermions hidden by vacuum structure of similar energy scale.  
These objects have maximum stable masses of sub-planetary scale and radii of micron to centimeter scale. We describe the surface gravity and tidal forces near these compact ultra dense objects, as pertinent to signatures of their collisions with visible  matter objects. 

\end{abstract}

\begin{keyword}
Dark Matter \sep Compact objects
\PACS 95.35.+d \sep 04.40.Dg \sep 12.39.Ba
\end{keyword}



\maketitle

\section{Introduction}

We present properties of gravitationally bound dark matter at energy scale 1-100 TeV. Our solutions lead to \underline{c}ompact \underline{u}ltra \underline{d}ense \underline{o}bjects (\CUDO s) with upper mass limit $M<0.1M_{\oplus}$, where $M_{\oplus}=6\times10^{27}\:{\rm g}$ is the Earth's mass.  This work thus implements recent interest in the mass scale 1-100 TeV for particle dark matter~\cite{Taoso:2007qk,Hooper:2009zm,Feng:2010gw}, building on prior investigations of gravitationally self-bound objects made of non-baryonic (dark) matter~\cite{Markov:1964,Gao:1980wc,Kolb:1993zz,Kolb:1994fi,Jin:1999we,Bilic:2001iv,Zurek:2006sy,Narain:2006kx,Dai:2009ik}.  

The factor thousand or more separation from baryonic matter means that the \CUDO~density scale is $10^{26}\:{\rm g/cm}^3$.  This high density implies that \CUDO s with upper limit (sub-planetary) masses $M<0.1M_{\oplus}$ have radii $R<10\:{\rm cm}$, making them presently impossible to detect by astronomical methods.  Here, note that the distribution of compact objects remains weakly constrained in this mass regime, see Fig.~8 of~\cite{Carr:2009jm}.  Dark matter \CUDO s are found to have strong gravity and sub-planetary mass, suggesting as observational signatures impacts on solar system bodies~\cite{Labun:2011wn}.  Our goal here is to provide a reference for investigating impact characteristics by evaluating \CUDO s' size and gravitational properties.

In our consideration of gravitationally-bound dark matter \CUDO s, the mass scale $1-100\:{\rm TeV}$ may represent either a) the mass of an isolated dark matter particle, or b) the energy scale of vacuum structure hiding the interactions of the dark matter sector.  Gravitational binding is computed for (case a) a degenerate fermi gas of TeV-mass fermions and for (case b) a bag-model equation of state with TeV-scale bag pressure.   We do not address \CUDO s of other types, such as composed of standard model particles~\cite{Zhitnitsky:2006vt,Cumberbatch:2006bj} or supersymmetric $Q$-balls.  $Q$-ball collisions with Earth are considered in~\cite{Kusenko:2009iz}, which provides an extensive list of references.

The \CUDO~types we consider are simple models of dark matter in and above 
the TeV-scale  paralleling the structure of the visible MeV-GeV-scale matter. Thus we consider both elementary dark-particles, and vacuum-struc\-tured dark-quark composite dark-had\-rons.  We discuss large dark-quark bags, though in the present view of visible matter the presence of quark matter in stars remains hypothetical~\cite{Weissenborn:2011qu}, and primordial quark nug\-gets are expected to evaporate~\cite{Madsen:1998uh}.  A different outcome can arise for dark matter when compared to similar theoretical structure of visible matter for differently fine-tuned parameter set. To see this, recall the fine-tuning of parameters leading to the properties of visible matter~\cite{Hogan:1999wh}.

Considering dark-matter that parallels visible matter but with different natural constants, we expect spectacularly different outcome for the properties of the dark Universe compared to the visible Universe. This is in fact suggested by the history of exploration of strange quark matter, that suggests a diversity of phenomenology of the dark-quark matter properties. The early studies of strange quark matter illustrate how quark matter phase structure can be sensitive to theory parameters, such as quark masses and the strong interaction QCD coupling constant.  When knowledge of these parameters was less established, choices could be made to render strange quark matter stable~\cite{Witten:1984,Farhi:1984qu}, leading to strange quark matter objects (strangelets) from femtometer to kilometer size~\cite{Madsen:1998uh,Weber:2004kj}.

Just like strangelets were stabilized in early models~\cite{Witten:1984,Farhi:1984qu}, dark-quark matter can be stabilized, either by (i)  a greater (speaking in relative terms) dark-quark-quark interaction strength,  and/or (ii) a smaller mass of dark-`strange'-quark. To be specific, instead of the  dark sector coupling $\alpha_d(\mu=100\mathrm{\,TeV})=0.117$ paralleling the QCD value at the higher scale, consider $\alpha_d(\mu=100\mathrm{\,TeV})=0.25$, twice as large.  In addition, consider the dark-strange quark half as massive compared to the scaled value, i.e.  $m_{ds}(2\mathrm{\,TeV})=45\mathrm{\,GeV}$.  Based on QCD experience, we believe that these choices will suffice to: (i) make the most stable object in the dark Universe a dark-Lambda, (ii) alter decisively dark-hadron interactions sensitive to dark chiral symmetry-breaking through the response of the dark pion (Goldstone boson) mass, akin to QCD case~\cite{Maris:1997hd,Chen:2003im,Bazavov:2009bb}, and (iii) stabilize dark-quark matter \CUDO s analogous to strangelets.

Considering prior QCD work~\cite{Witten:1984}, stronger dark-quark coupling can make dark-quark matter formation possible during transition to the dark-confining phase of the early Universe.  Many body effects, akin to color-flavor locking or particle-antiparticle locking~\cite{Alford:2007xm,Fukushima:2010bq}, introduce instabilities in the dark-quark phase near to dark sector confinement scale temperature in the early universe.  Such instabilities can promulgate large density fluctuations through the phase transition to become progenitors of the gravitationally self-bound objects we investigate.  As a consequence, dark-quark matter once formed can be stable, even when its mass is too small for gravity to play an important role.
 
Matter-antimatter symmetry in the dark sector is a potential objection to gravitationally bound dark stars based on QCD-like model of the dark-matter sector of the early Universe.  However, dark-quark-matter symmetric in matter and antimatter evaporates into, among others, the dark-Goldstone boson particles (the lowest mass dark-hadron) only if the dark-hadron gas is energetically favored over a dark-quark matter many-body ground state~\cite{Alford:2007xm,Fukushima:2010bq}.  Another possible remedy is a large asymmetry of dark-quark matter and antimatter, at percent-scale rather than $10^{-9}$ we observe among visible matter. In such a situation,  emission of dark-pions would rapidly cool the dark-quark matter, facilitating rapid gravitational structure development.

If the dark matter-antimatter asymmetry is not present by virtue of initial conditions of the Universe, developing the asymmetry requires the three Sakharov conditions~\cite{Sakharov:1967dj}.  Of these, there is only certainty about the irreversibility in the early, fast expanding Universe.  Dark-baryon number-violating processes and the strength of $C$- and $CP$-violation in the dark sector are unconstrained. As noted for the visible Universe~\cite{Hogan:1999wh}, the world of dark matter, even if identical in its theoretical structure to the visible Universe, will depend decisively on the values of parameters which govern it as well as the initial conditions of the Universe, as we now note.

This discussion shows that QCD/quark-matter experience invites the possibility of phenomenology leading to dark matter \CUDO s.  The practical absence of experimental constraints means that to make progress we must propose qualitative models and explore the observational constraints  that arise.  This is what we do in this manuscript with regard to exploration of gravitationally bound objects composed of dark matter. We take an approach inclusive of many possible microscopic models of the dark matter sector and contribute to constraining the model space by exploring observable characteristics of dark \CUDO s: In this spirit, in Sec.~\ref{sec:MRevals}, we quantify the relation between the dark matter energy scale and the \CUDO s' mass and radius.  We then discuss in Sec.~\ref{sec:graveffects} gravitational effects near their surfaces, and in Sec.~\ref{sec:cosmology} origins and cosmological impacts.  Section~\ref{sec:concl} concludes.

\section{Mass and Radius of CUDOs}\label{sec:MRevals}

\begin{figure*}[t]
 \includegraphics[width=0.49\textwidth]{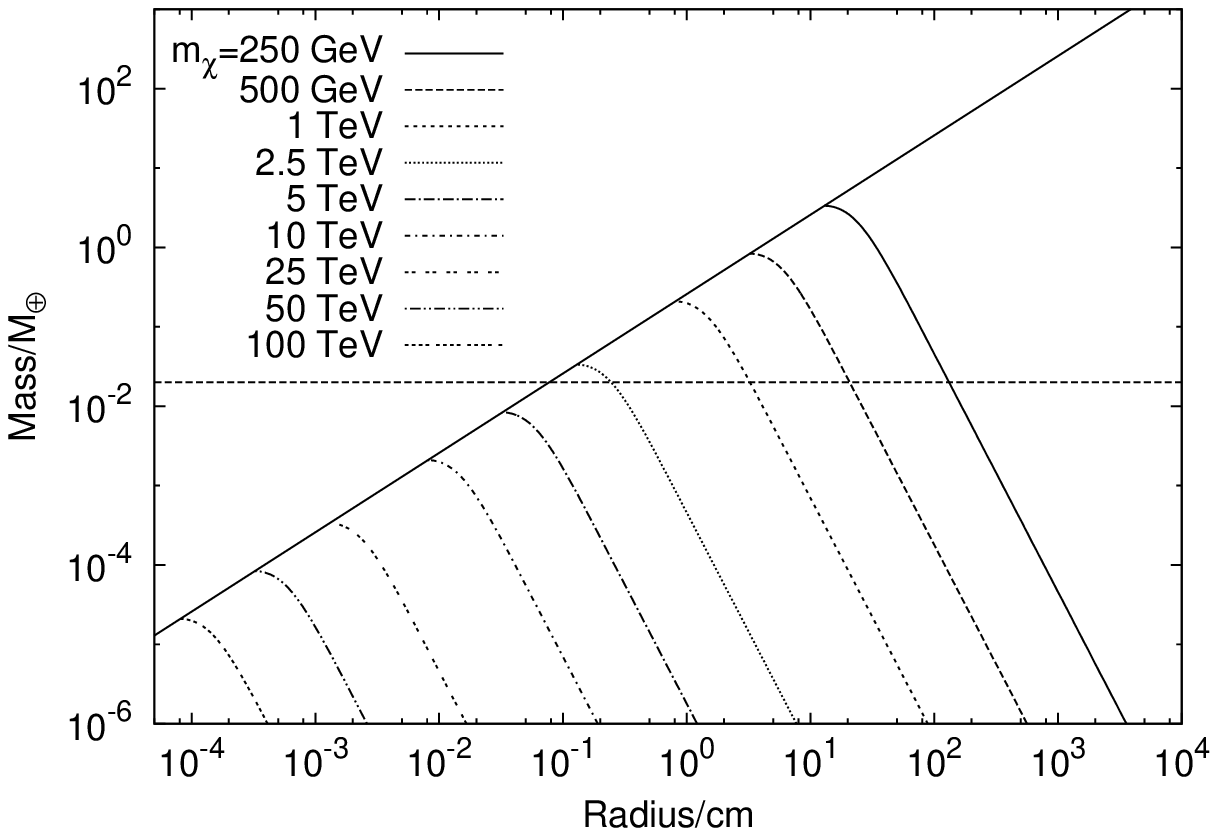}
 \hspace{1mm}
 \includegraphics[width=0.49\textwidth]{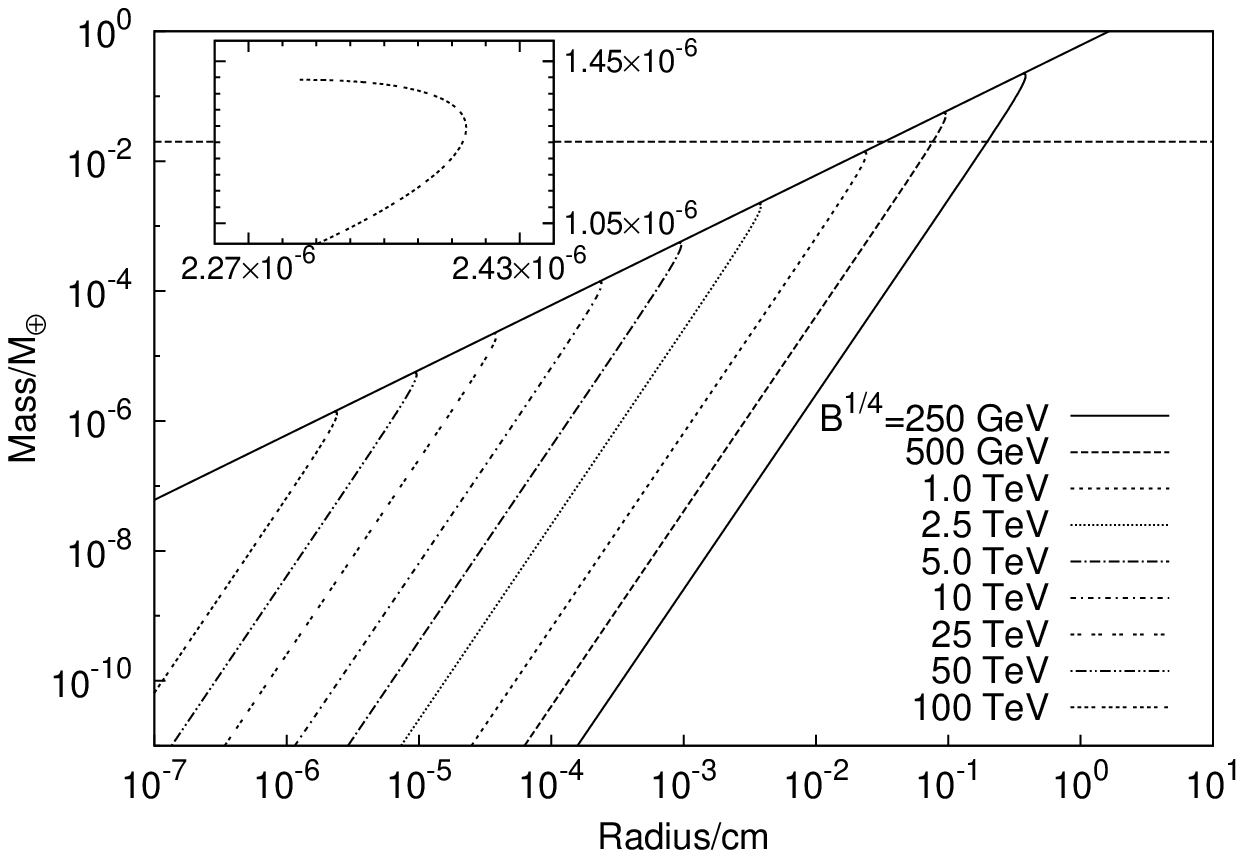}
\caption{\label{fig:massradius} Mass-radius relations obtained from the TOV equations for different fermion masses (left) or $B^{1/4}$ (right), with $g=2$ for both.  The continuous line shows the extremal configurations corresponding to Eqs.\:\eqref{fermiMR} and \eqref{vacstarMR}.  The dashed horizontal line shows the constraint from recent microlensing surveys.  The inset in the right plot magnifies the top of the $B^{1/4}=100\:{\rm TeV}$ curve using the same axes as the host plot.}
\end{figure*}

The gross physical properties of the compact objects accounting for their gravity are obtained from the Tolman-Oppenheimer-Volkoff (TOV) equations~\cite{Weinberg}.  In compact objects composed of massive particles supported by fermi-degeneracy pressure, the scale is determined by the critical density $\rho_c= m_{\chi}^4/3\pi^2$ where $m_{\chi}$ is the mass of the isolated dark matter particle in vacuum.  The extremal configuration is determined by $\rho_c^{-1/2}$~\cite{Narain:2006kx,Weinberg} to have maximum mass and corresponding (minimum) radius
\begin{subequations}\label{fermiMR}\begin{align}
\label{fermiM}
M_{\rm max} &= \frac{0.209}{(g/2)^{1/2}} \left(\frac{\rm 1\: TeV}{m_{\chi}}\right)^{2} M_{\oplus} \\
\label{fermiR}
R &= \frac{0.809}{(g/2)^{1/2}} \left(\frac{\rm 1\: TeV}{m_{\chi}}\right)^{2} \!\!\text{cm}=8.74\:GM_{\rm max},
\end{align}\end{subequations}
depending inversely on the square of the particle mass.  For comparison, the mass of Earth's moon is $1.2\%\:M_{\oplus}$ and a solar mass $M_{\odot}=2.0\times10^{33}\:{\rm g}=3.3\times10^5M_{\oplus}$.  

The numerical prefactor in \req{fermiMR} is obtained for the spin degeneracy of a single fermion $g=2$.  We can utilize $g$ to account for more than one species of fermion in the dark matter sector, provided their mutual interactions result in stable coexistence and they have nearly equal masses.  The $i$-th species contributes $g_i$ degrees of freedom to the total degeneracy factor $g=\sum{g_i}$.  From the fact that $\rho_{c}\propto gm_{\chi}^4$, we deduce the scaling of the extremal configuration \req{fermiMR}.  

The energy scale of structured-vacuum \CUDO s is set by the vacuum pressure (or `bag pressure') $B$.  Neglecting masses and interaction of constituent particles, the maximum mass of a structured-vacuum \CUDO~is~\cite{Lattimer:2006es}
\begin{subequations}\label{vacstarMR}\begin{align}
\label{vacstarM} 
 M_{\rm max} &= \frac{0.014}{(g/2)^{1/2}}\left(\frac{\rm 1\: TeV}{B^{1/4}}\right)^{\!2} M_{\oplus} \\
\label{vacstarR} 
 R &= \frac{0.023}{(g/2)^{1/2}}\left(\frac{\rm 1\: TeV}{B^{1/4}}\right)^{\!2}\:{\rm cm}
    = 3.69\:GM_{\rm max},
\end{align}\end{subequations}
inversely proportional to the bag pressure.  At the surface, the net pressure is zero.  Gravitational pressure is accounted for in the Schwarzschild ansatz for the line element and is thus implicit in the TOV equations.  The condition at the surface of the structured-vacuum \CUDO~is therefore that the vacuum pressure exactly balances the Fermi pressure $B=P_f$.  Because the Fermi pressure is directly proportional to $g$, we must have $B\propto g$.  The scaling of bag model stars in Sec. III.C of~\cite{Narain:2006kx} then result in the scaling of the extremal mass and radius in~\req{vacstarMR}.  Unlike for massive-fermion \CUDO s, the radius of the extremal struc\-tured-vacuum \CUDO~is neither a maximum nor minimum among the stable configurations.

Mass-radius relations obtained from numerical integration of the TOV equations are plotted in Figure~\ref{fig:massradius}.  It must be noted that determination of the radius is sensitive to numerical tolerances: the physical condition that pressure vanishes is verified only to within the numerical precision of the integration code.  As a result, computed radii vary slightly among evaluations, compare e.g.~\cite{Gao:1980wc,Narain:2006kx,Witten:1984}. Using the same integration technique as~\cite{Narain:2006kx} with $>10$ times smaller radius step size for high central densities, we determine the surface as the radius at which the pressure crosses zero and obtain a dimensionless radius $m_{\chi}^2\sqrt{G}R=3.357$ at the dimensionless central density $\rho(0)/m_{\chi}^4=0.02326$ for the extremal configuration of a massive-fermion \CUDO.  Our smaller step size suggests that these values agree with the finding of~\cite{Narain:2006kx} to within the numerical precision of the previous work. Correspondingly, we obtain a surface radius $\sqrt{BG}R=0.09546$ at $\rho(0)/4B=4.812$ for a struc\-tur\-ed-vacuum \CUDO~in the extremal configuration confirming the values of \cite{Witten:1984}.  Our results agree with the scaling relation and numerical results for smaller masses ($m_{\chi} \leq 100 \:{\rm GeV} $) in \cite{Narain:2006kx} and for $B^{1/4}=145\:{\rm MeV}$ in \cite{Witten:1984}.

The rising solid line in each frame presents the extremal config\-urations defined by Eqs.\:\eqref{fermiMR} and \eqref{vacstarMR}.  Configurations above this instability line are gravitationally unstable and collapse.  Moving away from the instability line along a curve of particular $m_{\chi}$ or $B$ corresponds to decreasing central energy density of the \CUDO.  The shape of the curves is for each type independent of $m_{\chi}$ or $B$. In particular, the power law behavior is due to the scaling solutions of the Lane-Emden equation for the non-relativistic equation of state and constant density star respectively for the massive-fermion and structured-vacuum \CUDO s~\cite{Narain:2006kx,Weinberg}.

\begin{figure*}
 \includegraphics[width=0.49\textwidth]{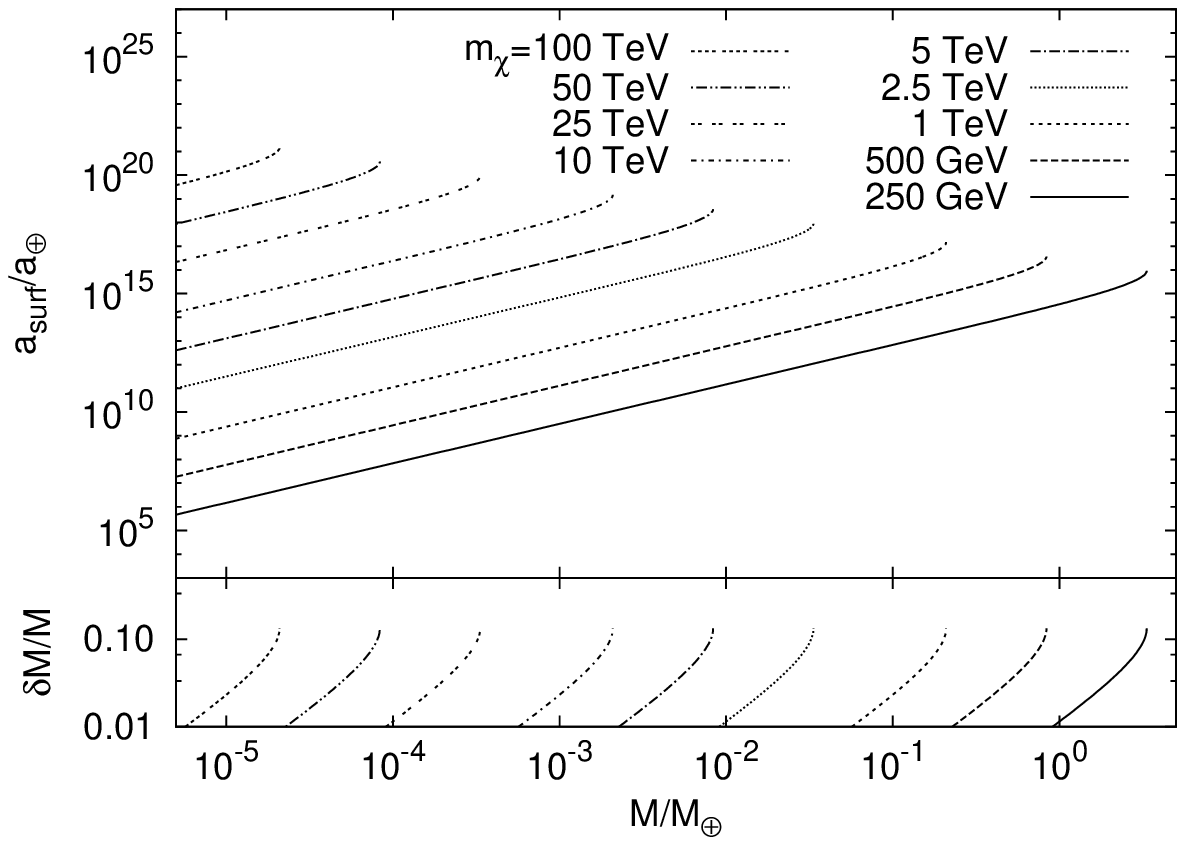}
 \hspace*{1mm}
 \includegraphics[width=0.49\textwidth]{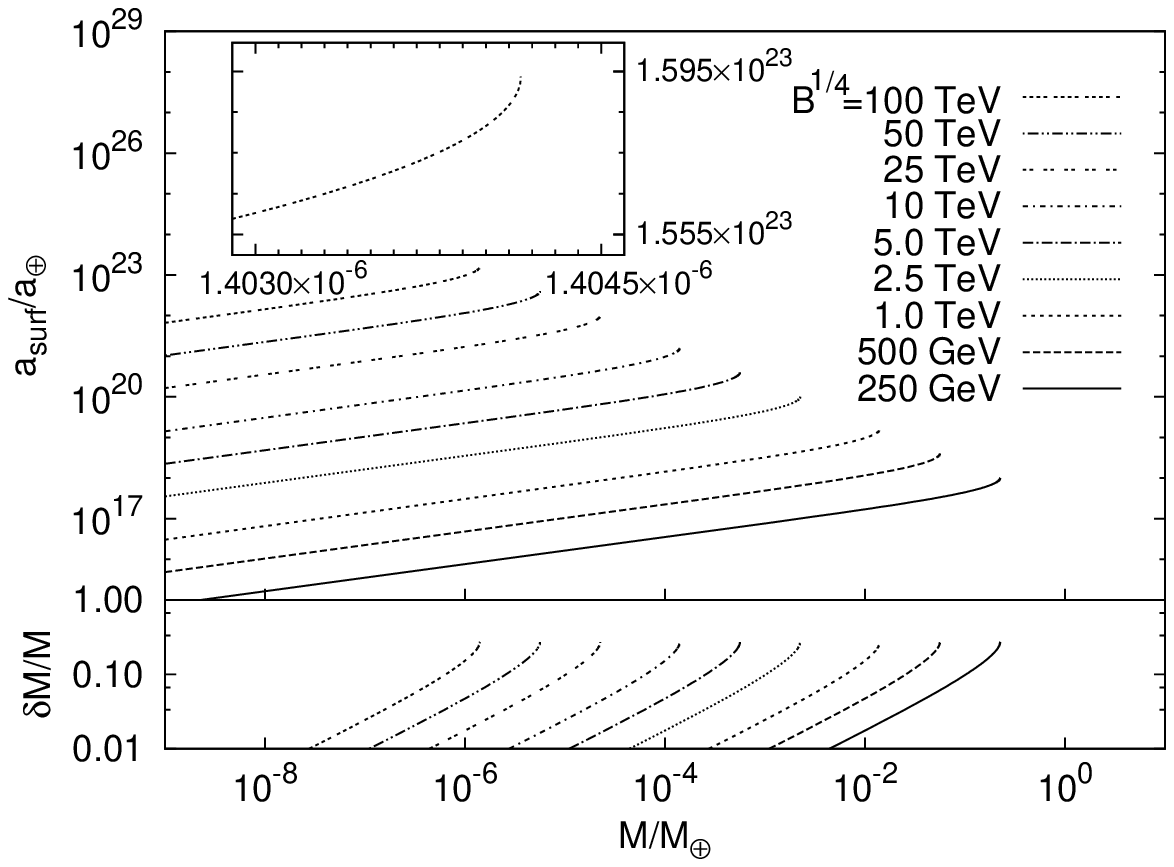}
\caption{\label{fig:gravity} Gravitational surface acceleration and mass defect (lower frame) for a massive-fermion \CUDO~(left) and structured-vacuum \CUDO~(right) as a function of total mass.  The surface acceleration and \CUDO~mass are scaled by their respective values for the Earth.  The inset and key are as in Fig.\ref{fig:massradius}.}  
\end{figure*}

For a massive-fermion \CUDO~composed of a single spe\-cies of fermion ($g=2$) with dark particle mass ranging from 250 GeV to 100 TeV, we find masses up to $1 M_{\oplus}$.  Gravitational microlensing surveys limit less than 20\% by mass of dark matter in the Milky Way's halo to be found in objects with mass $M>2\times 10^{-2}M_{\oplus}$~\cite{Tisserand:2006zx}.  The dotted horizontal line in Figure~\ref{fig:massradius} marks the microlensing limit and, assuming all dark matter is in \CUDO s, suggests a safe lower bound on the dark matter particle mass $m_{\chi}\gtrsim 3\:{\rm TeV}$.  Radii of \CUDO s then likely range up to a few cm.  For a structured-vacuum \CUDO, we obtain masses up to $0.1M_{\oplus}$ varying $B^{1/4}$ from 250 GeV to 100 TeV, which translates into a safe lower bound on the bag pressure $B^{1/4}\gtrsim 1\:{\rm TeV}$.  Since not all dark matter is in \CUDO s, these limits are soft.

\section{Gravitational Effects Near the Surface}\label{sec:graveffects}

For \CUDO s to survive close encounters with other astronomical objects, they must be stable against gravitational disturbance.  The surface acceleration $a_{\rm surf}$ is defined by the locally-measured force required to stop a particle at the surface from falling freely.  The radial component of the four-acceleration in the local frame of a stationary observer is $a^{r}=\Gamma^r_{tt}u^tu^t\sqrt{g_{rr}}$~\cite{MTW}.
Evaluated in standard Schwarzschild coordinates we obtain
\begin{equation}
\label{surfacc}
a_{\rm surf} = a_{\rm N} \left(1-\frac{2GM}{R_{\rm surf}}\right)^{-1/2}
\end{equation}
in which $a_{\rm N}=GM/R_{\rm surf}^2$ is the Newtonian surface acceleration. 
Using the extremal configuration of a structured-vacuum \CUDO, we find that the acceleration $a_{\rm N}$ is enhanced by $(1-2GM_{\rm max}/R)^{-1/2}=1.48$. For a massive-fermion \CUDO, the modification is smaller, $a_{\rm surf}=1.14a_{\rm N}$ for the extremal configuration.
The total modification due to general relativistic correction falls thus in the range
\beqn
1\leq \frac{a_{\rm surf}}{a_{\rm N}}\leq 1.48\:.
\eeqn
Because of the smallness of the correction, $a_{\rm surf}$ displayed in Fig.~\ref{fig:gravity} demonstrates the magnitude of $a_{\rm N}$ as well.

For a massive-fermion \CUDO~in the non-relativistic regime, characterized by Fermi momentum $k_F \ll m_{\chi}$, general relativistic corrections to Newtonian gravity are negligible. For these objects, the scaling solutions of the Lane-Emden-equation~\cite{Weinberg} can be used to write 
\begin{equation}\label{aNewt}
\frac{a_{\rm N}}{a_{\oplus}} = 
5.17\times10^{17}\left(\frac{m_{\chi}}{1\:\rm TeV}\right)^{\!16/3}\left(\frac{M}{M_{\oplus}}\right)^{\!5/3},
\end{equation}
in which $a_{\oplus}=9.81\:{\rm m/s}^2$ is the gravitational acceleration at the surface of the Earth. 

Near to the maximum mass for a given $m_{\chi}$ or $B$, general relativistic corrections to the surface gravity become relevant.  One considers the mass defect
\beqn\label{massdefect}
\frac{\delta M}{M} = \frac{1}{M}\int_0^{R_{\rm surf}} \!\!\rho\, 4\pi r^2(\sqrt{g_{rr}}-1)dr.
\eeqn 
Figure 2 displays the surface acceleration \req{surfacc} and mass defect \req{massdefect} of the configurations shown in Fig.\:\ref{fig:massradius}. For a massive-fermion \CUDO, one finds a maximum mass defect of $13 \%$.  The greater compactness of structured-vacuum \CUDO s implies that a maximum mass defect of $27\%$ is possible.  These maximal mass defects are independent of $m_{\chi}$ or respectively $B$, since the scaling of the integrand in Eq.~\ref{massdefect} and $M$ is identical.

If a Hawking temperature is a general characteristic of strong gravitational fields (as suggested by e.g.~\cite{Barcelo:2010pj}), the accelerations seen in Figure\:\ref{fig:gravity} correspond to Hawking temperatures
\beqn
T_H = \frac{a_{\rm N}}{2\pi} 
 = 3.98\times10^{-20}\:\frac{a_{\rm N}}{a_{\oplus}}\:{\rm K}
 = 3.43\times10^{-24}\:\frac{a_{\rm N}}{a_{\oplus}}\:{\rm eV}.
\eeqn
Extremely small surface area makes the total luminosity too small ($< 10^{-7}\:{\rm J/s}$) to be observable, but the low rate of energy loss also means that \CUDO s have evaporation lifetimes well beyond the age of the universe.

The gravitational tidal acceleration near the surface of the \CUDO~is
\beqn
a_{\rm tidal}=\frac{2GML}{R^3}
 =a_{\rm N}\frac{R_{\rm surf}^2}{R^2}\frac{2L}{R},
\eeqn
where $L$ is the length scale over which the tidal force acts.  Normal density matter behaves as a collection of individual particles when the tidal potential $\Delta V_{\rm tidal}= ma_{\rm N}LR_{\rm surf}^2/R^2$ on neighboring nuclei, $m\sim 28 m_p$ for silicon, exceeds the average interatomic potential $V\sim 1\:{\rm eV}$ over the interatomic distance $L\sim 1\:$\AA.  The critical surface acceleration $a_{\rm c}$ at which the tidal potential overcomes the interatomic binding is $a_{\rm c}=3.5\times10^{15}a_{\oplus}$. Since $a_{\rm surf}\sim a_{\rm N}$, Fig.~\ref{fig:gravity} shows that $a_{\rm c}$ can be exceeded even by \CUDO~configurations with $m_{\chi}=250\:{\rm GeV}$ or $B^{1/4}=250\:{\rm GeV}$.

Although $a_{\rm tidal}$ decreases as $R^{-3}$, we see that normal density matter up to distances many times the surface radius cannot resist the passage of a \CUDO.  The interaction of the \CUDO~with the normal matter medium is therefore primarily in the deposition of energy via atomization of the medium by tidal forces.

\section{Primordial Origin and Impact on Cosmology}\label{sec:cosmology} 

We follow the hypothesis that \CUDO s originate in the early universe, prior to big bang nucleosynthesis.  We are led to consider dark matter particles with masses above 3 TeV gravitationally unstable by four factors:
\\ {\bf a)} Becoming non-relativistic at an earlier time, dark matter has a density proportionally higher at the time when gravity can begin to work on local density fluctuations~\cite{Kolb:1994fi}.  
\\ {\bf b)}  The final compact object comprises $10^{11}-10^{19}$ fewer particles (increasing with $m_{\chi}\geq 3\:{\rm TeV}$), requiring therefore a smaller initial volume to collapse. 
\\ {\bf c)}  The gravitational dark particle-dark particle interaction strength is $10^6-10^{10}$ times larger.
\\ {\bf d)}  Normal (visible) matter sharing the incipient gravity well is more easily ejected and carries energy and angular momentum away from the collapsing dark matter.  \\
We expect massive-fermion \CUDO s can coalesce rapidly, with a time\-scale for collapse accelerated compared to baryonic stars.  Numerical determination of the collapse time\-scale would be provided by detailed modeling of the dynamics of high mass dark matter particles in the rapidly expanding and cooling background of the universe.

The density of free-streaming dark matter particles is reduced by sequestering a fraction of dark matter in compact ultra dense objects.  Direct detection experiments on Earth then have a more challenging task due to greater diluteness of the residual dark matter particle gas.  We emphasize that these qualitative considerations are consistent with the prevailing view that dark matter with a lower mass-energy scale remains a gas of single particles. 

Structured-vacuum \CUDO s are stabilized by vacuum pressure resulting from a phase transition to the confining vacuum at a temperature of the order of magnitude of $B$.  During the phase transition, some fraction of dark matter is confined to residual bubbles of the high temperature phase that could persist into the present era as structured-vacuum \CUDO s.  

The remaining fraction of dark matter emerges as free-streaming single particles in the low temperature phase.  These particles have mass (or masses) estimated from the bag model~\cite{Johnson:1975}
\beqn \label{mchibag}
m_{\chi} \lesssim \frac{4}{3}(4\pi B)^{1/4}(2.04 g)^{3/4} .
\eeqn
Recall that $g$ is the total number of massless degrees of freedom propagating in the false vacuum of the interior of the object.  The inequality is due to the fact that only a subset $N_c\leq g$ are charged under the confining gauge group and contribute to a (lowest) singlet state.  Combining lower limits on $m_{\chi}$ with the lower limits on $B$ implied by our results assists in constraining the number of degrees of freedom in the dark sector.

Dark matter \CUDO s, having primarily gravitational interaction with visible matter, are consistent with big bang nucleosynthesis.  Nucleons within a nucleus near a \CUDO~are accelerated apart when $\Delta V_{\rm tidal}\gtrsim 1\:{\rm MeV}$ and $L=1\:{\rm fm}$ corresponding to $a_c\sim 10^{27}a_{\oplus}$, showing that tidal accelerations near \CUDO s are insufficient to influence nuclei.  We note also that the maximal number density of \CUDO s (comprising all dark matter) at the time 
\beqn
\frac{N}{V} = 3\times10^{-27}\frac{M_{\oplus}}{M} {\rm cm}^{-3}
\eeqn
is 47 orders of magnitude (for $M=10^{-4}M_{\oplus}$) smaller than the number density of baryons.  \CUDO s' influence over density distribution and reaction rates during nucleosynthesis is expected to be negligible.

Assuming gravitational interactions dominate\ \CUDO s' interaction with each other, \CUDO s are consistent with constraints on the scattering cross-section of dark matter implied by structure formation and present-day observations.  The cross-section for gravitational scattering scales as~\cite{Padmanabhan}
\beqn
\frac{\sigma_G}{M} \approx \frac{1}{M}\left(\frac{GM}{v^2}\right)^{\!2}
=3\times10^{-15} \left(\frac{\rm 100\: km/s}{v}\right)^{\!4}\frac{M}{M_{\oplus}}\:\frac{\rm cm^2}{g}
\eeqn  
which is well within the bounds set by the Bullet cluster, $\sigma/m <0.7\:{\rm cm^2\:g^{-1}}$~\cite{Markevitch:2004,Randall:2008}.

\CUDO s provide natural seeds for subsequent small-scale structure formation and should accelerate the collapse and lifecycle of the first stars.  `Free' \CUDO s, not at the center of a normal matter star, could then be a result of ejection during violent astrophysical events~\cite{Bauswein:2008gx}.  

Conversely, any normal matter associated with the free \CUDO~(e.g. trapped in its gravitational potential well) has negligible effect on its externally observed characteristics.  The presence of normal matter dressing the \CUDO~affects their identification by astrophysical or planetary~\cite{Labun:2011wn} signatures.

\section{Conclusions}\label{sec:concl}
The energy scale $>3\:{\rm TeV}$ is outside the discovery potential of particle physics experiment for many years to come.  Our detailed results for the typical mass and size scales of dark \CUDO s motivate efforts to refine astrophysical and planetary observations in search of constraints on the parameters of the dark matter sector.  Stronger constraints on the mass-energy scale of dark matter would be provided by microlensing studies sensitive to objects in the sub-planetary ($M\lesssim M_{\oplus}$) mass range.  On the other hand, \CUDO s' unhindered passage through normal matter invites searches for signals of gravitational interactions of \CUDO s with visible matter bodies~\cite{Labun:2011wn}.

We have investigated the properties of compact objects composed of (1) massive fermions and (2) massless fermions confined by beyond-the-standard-model vacuum structure.  Constraints we derive from astrophysical considerations are independent of theoretical challenges in formulating models of particle dark matter.  Our results are inclusive of most dark matter theories without attempting to determine in detail how any specific variant can (or cannot) be realized.  More work will be necessary to understand issues of baryogenesis of the TeV-mass fermions, phase structure of the hidden vacuum sector, and/or clustering dynamics. 

Numerical evaluations of the mass and radius of stable objects are presented in Figure\:\ref{fig:massradius}.  We have discussed consequences of \CUDO s' characteristically strong surface gravities, which are demonstrated in  Figure\:\ref{fig:gravity} to range up to $10^{23}$ times that of Earth.  We find that such compact objects have $\mu{\rm eV}$-meV Hawking temperatures, and tidal forces in their vicinity are large enough to atomize normal density matter.  

Our results show that whether dark matter involves an elementary fermion of TeV mass or vacuum structure of TeV scale, compact objects formed of dark matter have maximum mass less than the Earth's.  Combined with limits on masses of compact objects in the galactic halo placed by microlensing surveys~\cite{Tisserand:2006zx}, our results suggest a safe lower limit on particle mass $3\:{\rm TeV}\lesssim m_{\chi}$ or energy scale of the dark matter sector vacuum structure $1\:{\rm TeV} \lesssim B^{1/4}$.   The higher mass-energy scale suggests dark matter \CUDO s have a primordial origin, which is consistent with standard cosmology.

\section*{Acknowledgments}
This work was supported by the grant from the U.S. Department of Energy, DE-FG02-04ER41318.


\end{document}